\documentclass{aa}  
\RequirePackage{etex} 
\usepackage{graphicx}
\usepackage{booktabs,multirow, threeparttable}
\usepackage{comment}

\usepackage{txfonts}
\usepackage{xcolor}
\usepackage{scalerel}
\usepackage{tikz}
\usetikzlibrary{svg.path}

\definecolor{orcidlogocol}{HTML}{A6CE39}
\tikzset{
  orcidlogo/.pic={
    \fill[orcidlogocol] svg{M256,128c0,70.7-57.3,128-128,128C57.3,256,0,198.7,0,128C0,57.3,57.3,0,128,0C198.7,0,256,57.3,256,128z};
    \fill[white] svg{M86.3,186.2H70.9V79.1h15.4v48.4V186.2z}
                 svg{M108.9,79.1h41.6c39.6,0,57,28.3,57,53.6c0,27.5-21.5,53.6-56.8,53.6h-41.8V79.1z M124.3,172.4h24.5c34.9,0,42.9-26.5,42.9-39.7c0-21.5-13.7-39.7-43.7-39.7h-23.7V172.4z}
                 svg{M88.7,56.8c0,5.5-4.5,10.1-10.1,10.1c-5.6,0-10.1-4.6-10.1-10.1c0-5.6,4.5-10.1,10.1-10.1C84.2,46.7,88.7,51.3,88.7,56.8z};
  }
}

\newcommand\orcidicon[1]{\href{https://orcid.org/#1}{\mbox{\scalerel*{
\begin{tikzpicture}[yscale=-1,transform shape]
\pic{orcidlogo};
\end{tikzpicture}
}{|}}}}
\usepackage{hyperref}
\usepackage{placeins}
\usepackage{lscape}

\definecolor{f}{rgb}{1., 0.44, 0.37}

\newcommand{\redd}{$R_{\rm Edd}$}
\newcommand{\mbh}{$M_{\rm BH}$}

\authorrunning{Aliaga et al.}
\titlerunning{Evidence for systematically suppressed fluctuations in the broad absorption line population}

\begin{document}

   \title{Continuum variability in AGN: Evidence for systematically suppressed fluctuations in the broad absorption line population}
   
   \subtitle{}

   \author{A. Aliaga 
          \inst{1,2,\orcidicon{0009-0008-0906-0057}}, P. Arévalo\inst{1,2,\orcidicon{0000-0001-5675-6323}}, M.L. Martínez-Aldama\inst{3,2,\orcidicon{0000-0002-7843-7689}}, W. Zuo \inst{4,\orcidicon{0000-0002-4521-6281}},
          B. Muñoz-Bravo \inst{5,2,\orcidicon{0009-0008-4374-6415}}
          }

   \institute{Instituto de Física y Astronomía, Universidad de Valparaíso, Gran Bretaña 1111, Valparaíso, Chile\\
              \email{aldo.aliaga@postgrado.uv.cl}
              \and Millennium Nucleus on Transversal Research and Technology to Explore Supermassive Black Holes (TITANS), Chile
              \and Astronomy Department, Universidad de Concepción, Barrio Universitario s/n, Concepción, 4030000, Chile
              \and Shanghai Astronomical Observatory, Chinese Academy of Sciences, 80 Nandan Road, Shanghai 200030, People’s Republic of China
              \and Departamento de Astronomía, Universidad de Chile, Casilla 36D, Santiago, Chile
}

   \date{}

% \abstract{}{}{}{}{} 
% 5 {} token are mandatory
 
  \abstract
   {}
   {We aim to compare the flux variability of broad absorption line quasars (BAL) to non-BAL quasars using excess variance and damped random walk (DRW) metrics, controlling for intrinsic differences in the black hole parameters of their populations to inform models concerning the generation of the BAL phenomenon.}
   {We constructed a sample of light curves of BAL and non-BAL quasars by cross-matching spectroscopically confirmed quasars from the SDSS DR16 catalogue to the Zwicky Transient Facility database, in the $g$ and $r$ bands, selecting only objects in the ranges of $1.57 \leq z \leq 2.00$ and $18.5 \leq  $ rmag $\leq19.8$. The redshift range ensures that the $g$ band covers the C\,\textsc{iv} emission line (and trough in BALs) while the $r$ band does not, mainly probing the continuum variability, and that the black hole masses (\mbh) can be estimated from the width of the Mg\,\textsc{ii} line for all objects.
   For each object, we quantified variability in both bands using the excess variance alongside the amplitude ($\sigma_{\mathrm{DRW}}$) and characteristic timescale ($\tau_{\mathrm{DRW}}$) of a DRW model. We also compared the DRW metrics in bins of \mbh \ and Eddington ratio (\redd) to isolate the influence of the BAL phenomenon in objects with the same physical properties.   }
   {The normalized excess variance and the $\sigma_{\mathrm{DRW}}$ are consistently smaller for BALs, confirming that BAL quasars display lower long-term variability; furthermore, the $g$ band exhibits higher variability than the $r$ band across both populations. These differences persist when the samples are compared within fixed bins of \mbh\ and \redd\, indicating that the suppressed variability in BAL quasars is not simply driven by differences in these properties between BAL and non-BAL samples.
   }
   {These results show that BAL quasars are systematically less variable than non-BAL 
quasars in both bands, confirming that this suppression is an intrinsic feature 
of the continuum rather than an effect of emission or absorption line contamination. The samples could be made to agree if the \mbh \ in the BAL sample were significantly overestimated, on average by a factor of $\gtrsim 4$. In this case their lower variability would be explained by their significantly higher \redd\ compared to the non-BAL sample. Alternatively, the lower variability in BALs can be related to their lower X-ray luminosities, or to significant nuclear obscuration, if the inner part of the accretion disc were more obscured in BALs and more variable than the rest of the disc.}


   \maketitle
%
%-------------------------------------------------------------------

\section{Introduction}

Active galactic nuclei (AGNs) often show signatures of energetic outflows driven by accretion onto supermassive black holes (SMBHs) that serve as potential feedback processes that contribute to the co-evolution of SMBHs and their host galaxies \citep{Ferrarese2000,Gulteltkin2009,Shen2015}
In quasars, these outflows are frequently traced by high-velocity absorbing gas along the line of sight, observed as blue-shifted absorption features in the UV spectrum. Approximately 20 per cent of AGNs have these characteristics and are classified as broad absorption line (BAL) quasars, depending on sample selection and wavelength coverage \citep{Bischetti2023}, exhibiting deep and wide absorption troughs of Si\,\textsc{iv}, C\,\textsc{iv} and N\,\textsc{v} (occasionally of Fe\,\textsc{ii} and Mg\,\textsc{ii}) \citep{Wampler1995}, observed blueshifted up to $\sim$0.2c \citep{rodriguez2020}
and with full widths at half-maximum exceeding $2000\ \mathrm{km\ s^{-1}}$ \citep{Weymann1991,hall2013}. Such broad blueshifted signatures provide direct evidence of powerful radiatively driven winds emerging from the accretion disc.

Optical and UV variability is a defining property of AGNs and has long been recognized as a powerful probe of the structure and dynamics of their central engine. As reviewed by \citet{Cacket2021}, variability offers one of the few indirect methods to study unresolved regions. Numerous studies have shown that at least part of the optical continuum variability arises directly from the accretion disc itself, rather than being uniquely a reprocessed response to rapid X-ray emission variations. This conclusion is supported by both short- and long-timescale analyses \citep[e.g.][]{Krolik1991,Arevalo2008,Ai2010,Edelson2015,Lira2015L,Smith2018}, firmly establishing optical variability as a key diagnostic of accretion physics. In BALs, these variability signatures may be further shaped by the presence of high-velocity absorbing gas. The wind can modulate the emergent UV/optical emission through changes in ionization \citep[e.g.][]{Filiz2013,Wang2015,He2017}, covering fraction \citep[e.g.][]{Gibson2008,Hamann2008,Hall2011,Filiz2012,Vivek2012,Vivek2016}, or shielding \citep[e.g.][]{Murray1995,Proga2000}, potentially imprinting specific variability patterns as compared to non-BAL quasars.
Given this connection, comparing the time-domain behaviour of BAL and non-BAL quasars offers a valuable way to probe whether the presence of powerful outflows is linked to UV/optical variability.

These BAL winds are often interpreted either as line-of-sight manifestations of a ubiquitous equatorial disk wind \citep[e.g.][]{Murray1995,Elvis2000,Giustini2019}, or as signatures of a short-lived evolutionary phase associated with rapid SMBH growth and feedback \citep[e.g.][]{Sanders1988,2007ApJ...667..149F,VillarMartin2020}. 
Therefore by comparing the variability between these two samples we can show how variability properties depend on viewing angle in the first scenario or on evolutionary stage, in the second. 

Most studies of BAL quasars have focused on absorption-line variability properties such as equivalent width (EW), trough depth, and velocity structure, and through their relation to continuum variability. A recurrent result is the anti-correlation between absorption-line strength and continuum flux, consistent with models in which changes in the ionizing continuum modulate the ionization state of the outflowing gas \citep[e.g.][]{VillarMartin2020,Mishra2021,Aromal2023,Qin2024}, reinforcing an ionization-driven scenario as the dominant mechanism. Consequently, variability studies of BAL quasars have largely emphasized line-driven diagnostics, with less emphasis on the statistical characterization of their continuum variability.

The stochastic nature of AGN variability requires the use of large samples. Variations observed in a single object may arise from random fluctuations rather than any real physical differences, so ensemble statistics allow us to isolate genuine differences between BAL and non-BAL quasars.
In this context, direct comparisons of continuum variability between BAL and non-BAL quasars remain scarce. To our knowledge, the only large-sample study addressing this question is that of \citet{Vivek2019}, which analysed Stripe~82 sources using the SDSS DR12Q variability parameters VAR\_A and VAR\_GAMMA, as defined in the DR12Q catalogue \citep{Paris2017}. \citet{Vivek2019} reports no statistically significant difference in continuum variability between the two populations. Because these metrics summarize variability through model parameters rather than direct variance estimates, complementary analyses employing excess variance and timescale-dependent variance estimators can help to further constrain potential differences between BAL and non-BAL populations.

In this work, we revisit this problem using a large sample from SDSS-IV DR16 Quasar catalogue (DR16Q; Lyke et al. \citeyear{Lyke2020}) of BAL and non-BAL quasars with multi-epoch optical photometry from  the Zwicky Transient Facility (ZTF; Masci et al. \citeyear{Masci2019}) Data release 23. For the redshift range considered here ($1.57 < z < 2.00$), the ZTF $g$ band probes the spectral region containing the \ion{C}{iv} BAL, while the $r$ band predominantly traces the adjacent UV–optical continuum.

This paper is organized as follows. Section~2 describes the construction of our sample, including the BAL classifications adopted from the SDSS DR16 catalogue and the criteria used to select well-sampled ZTF light curves. We outline the additional constraints applied to ensure a consistent dataset, such as limits on brightness and the restricted redshift interval adopted to place both BAL and non-BAL quasars at comparable rest-frame wavelengths.
Section~3 details the variability metrics employed in this study. We summarize the computation of the normalized excess variance and the damped random walk (DRW) model parameters, which allow us to characterize variability on long-term timescales, as well as to perform controlled comparisons across 
objects with similar \mbh\ and \redd.
Section~4 presents the statistical results, including population-level differences in variability between BAL and non-BAL quasars across both filters and at fixed \mbh\ and \redd.
Section 5 discusses the implications of these findings and explores possible explanations for the observed differences in variability, and in Section 6 summarizes our main conclusions.

\section{Sample and data}

Our parent sample is the SDSS-IV DR16 Quasar catalogue (DR16Q; Lyke et al. \citeyear{Lyke2020}), which contains over 750,000 spectroscopically confirmed quasars, and which includes ‘balnicity characteristics’. We selected quasars with \texttt{PBAL} = 0 and \texttt{PBAL} = 1. 
In \citet{Lyke2020}, \texttt{PBAL} = 0 corresponds to non-BAL quasars, while \texttt{PBAL} = 1 designates 
quasars whose C\,\textsc{iv} balnicity index (BI) \citep{Weymann1991} exceeds ten times its measurement uncertainty, 
allowing for a clean separation between BAL and non-BAL objects. We further restricted the sample to objects in the redshift range $1.57 < z < 2.00$, to allow a \mbh\ estimate from the Mg\,\textsc{ii} 2900 \AA\ broad line, and with a spectral signal-to-noise ratio of > 10, resulting in a sample of 13,994 quasars.

For each object in our final DR16Q sample, we attempted to retrieve optical light curves from the Zwicky Transient Facility (ZTF; Masci et al. \citeyear{Masci2019}) Data Release 23. ZTF is a wide-field optical time-domain survey conducted with the 48-inch Samuel Oschin Telescope at Palomar Observatory, using a 47~deg$^{2}$ focal plane camera that repeatedly scans the northern sky in $g$, $r$, and $i$ bands. We queried the ZTF public data  for sources centred on the SDSS coordinates of our sample and extracted multi-epoch $g$ and $r$-band photometry, obtaining light curves for 13,345 objects in the $g$ band and for 13,489 objects in the $r$ band.
These light curves span the period from March 2018 to October 2024 with a typical cadence of $\sim4$ days; however, coverage is non-uniform, as seasonal gaps and survey priorities cause some areas of the sky to be observed more often than others. We retained only good observing epochs by requiring the ZTF labels \texttt{CATFLAGS}\ = 0 and \texttt{LIMITMAG}\ > 20. We also removed multiple observations on the same night to produce more homogeneous light curves, by ensuring at least a 0.5 day separation between epochs. Since a given object can have observations with different field or ccd combinations, which overlap in time but can have small calibration offsets, we selected epochs only from the field or ccd combinations that had the largest number of points. We removed objects whose final light curves have less than 90 good epochs. After applying all these filters we retain 11,284 objects with light curves in $g$ and 11,304 with light curves in $r$.  Additionally, all 
light curves were shifted to the rest-frame to prevent cosmological time dilation 
from artificially inflating the characteristic variability timescales (e.g. $\tau_{\mathrm{DRW}}$), 
by dividing all the observing times by $(1+z)$, where $z$ is the source redshift.

\begin{figure}
    \centering
    \includegraphics[width=0.9\linewidth]{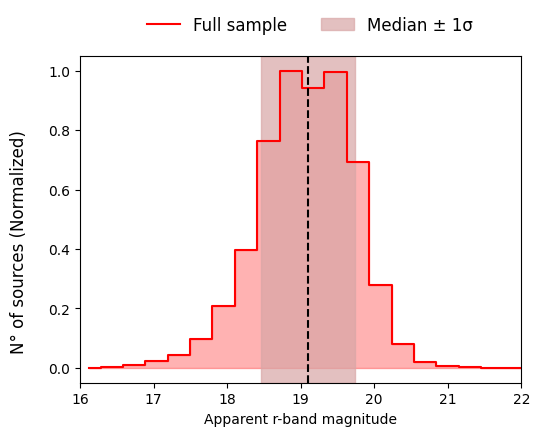}
    \caption{Distribution of magnitudes of the sample. The dashed orange line represents the median of the magnitudes and the grey area encloses the magnitudes within the median $\pm 1\sigma$}
    \label{fig:mags}
\end{figure}

This ensemble of light curves provides time-domain information to investigate the variability properties of BAL and non-BAL quasars identified in the SDSS DR16Q catalogue. Based on the $r$-band magnitude distribution shown in Fig. \ref{fig:mags}, we selected only sources with $r$-band magnitudes lying within the grey area (i.e. with $r=19.10 \pm 0.65$) so that BAL and non-BAL samples share the same narrow brightness range.
In addition, we restricted the analysis to quasars with $\log M_{\mathrm{BH}}$ and $\log R_{\mathrm{Edd}}$ values in ranges from $7.0$ to $10.5$ and -$2.0$ to $1.0$, respectively, which correspond to the most representative regions of the distributions of these parameters in our sample.

After applying these selection criteria, we retained only the 7,655 objects that have valid light curves in both bands. The final light curves used to compute the variances  have a median of 365 and 382 epochs in the g and r bands, with a mean
absolute deviation (MAD) of 92 and 90, respectively. When separating the sample into 
non-BAL ($n = 6624$) and BAL ($n = 1031$) groups, the sampling statistics 
remain highly consistent between the two populations. In the $g$ band, the non-BAL 
sub-sample has a median of 363 epochs ($\text{MAD} = 91$) compared to 368 epochs 
($\text{MAD} = 92$) for the BAL sample. Similarly, in the $r$ band, the non-BAL 
group shows a median of 382 epochs ($\text{MAD} = 90$), while the BAL group 
has a median of 389 epochs ($\text{MAD} = 94$). This close agreement in both the 
number of epochs and baseline lengths ensures that both light curve populations are  
equivalent, thereby mitigating potential biases in the variability statistics of each sample 
in our subsequent analysis.

To estimate the accretion properties of our sample, we computed \redd\ using bolometric luminosities ($L_{\mathrm{bol}}$) derived from the 3000\,\AA\ continuum luminosity from \citet{Wu2022}. We combined these with single-epoch \mbh\ estimates from the same catalogue, which were derived using the Mg\,\textsc{ii} emission-line measurements reported in the SDSS DR16 quasar catalogue \citep{Lyke2020}. We calculated the Eddington ratio as 
$R_{Edd} \equiv L_{bol}/L_{Edd}$, where $L_{Edd}=1.3 \times 10^{38}$\mbh$\, erg\, s^{-1}$.

\begin{figure*}[!h]
    \centering
    \includegraphics[width=0.9\linewidth]{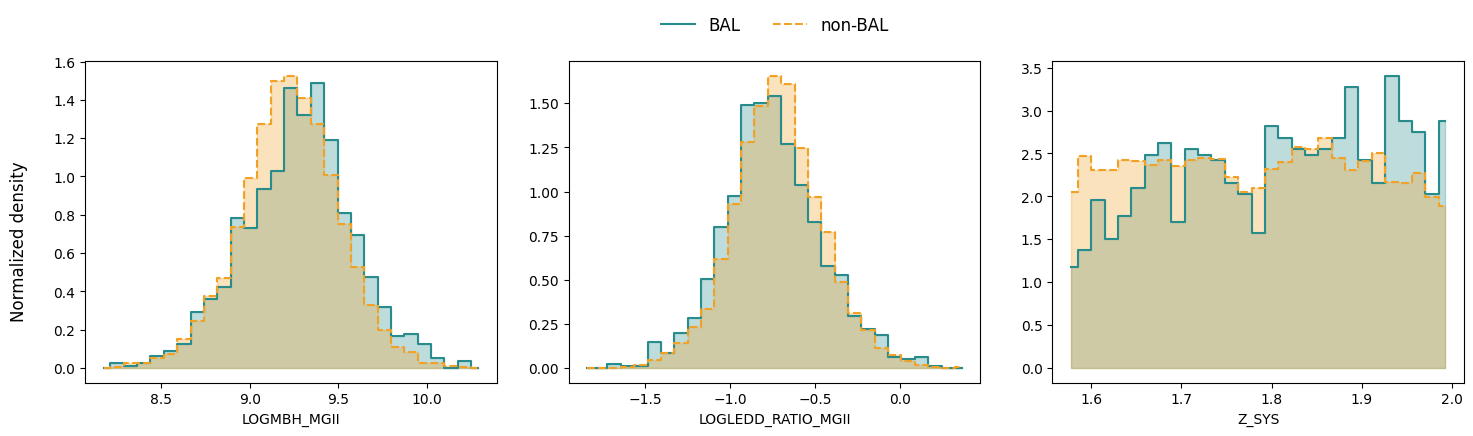}
    \caption{Sample property distributions for BAL and non-BAL quasars. From left to right, the panels show the black hole mass \mbh(Mg\,\textsc{ii}), the Eddington ratio \redd\ derived from \mbh(Mg\,\textsc{ii}), and the redshift corrected for systematics. Solid histograms represent the BAL sample, while dashed histograms show the non-BAL sample.}
    \label{fig:properties}
\end{figure*}

\citet{Wu2022} report \redd\ based on Mg\,\textsc{ii} for objects with $z<1.9$. Therefore, this calculation was necessary to ensure that the whole redshift range (i.e. also the objects with $1.9<z<2$) had consistent estimates. Fig. \ref{fig:properties} shows the distributions of \mbh, \redd\ and the systematics-corrected redshift estimated by \citet{Wu2022}, for the BAL and non-BAL samples. 

\begin{figure}
    \centering
    \includegraphics[width=0.9\linewidth]{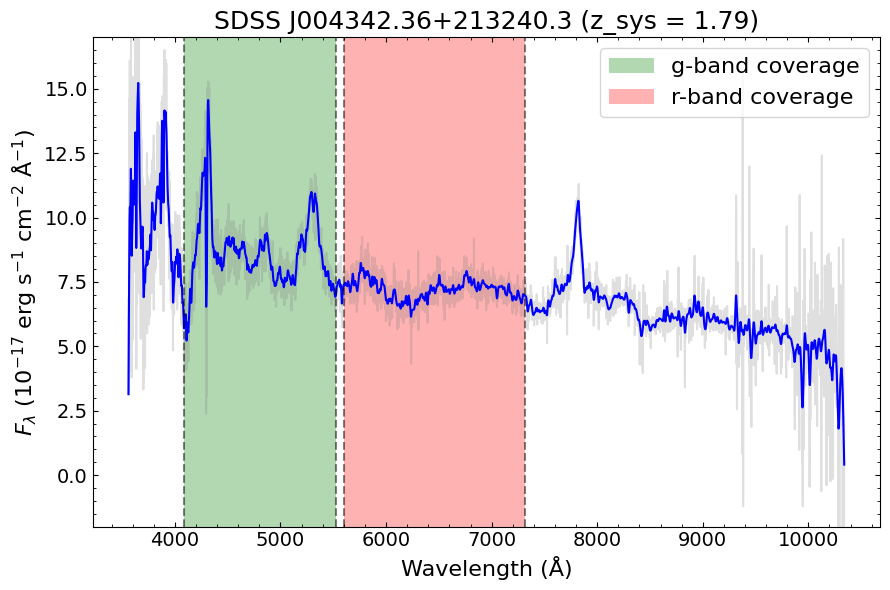}
    \caption{Example spectrum of a BAL quasar from the sample, at an intermediate redshift. The blue and red patches indicate the wavelength coverage of the ZTF $g$ and $r$ bands. At this redshift, the $g$ band coincides with the C\,\textsc{iv} emission line and the associated broad absorption region, while the $r$ band covers mostly continuum.}
    \label{fig:spec_bal}
\end{figure}

In Fig.~\ref{fig:spec_bal} we show the SDSS spectrum of one of the objects in the BAL sample, at redshift $z\sim1.8$, i.e. close to the centre of the redshift distribution. The shaded areas show the transmission windows of the $g$ filter, in blue, and the $r$ filter, in red. At these redshifts, therefore, the $g$ band observes the strong emission lines C\,\textsc{iii}] 1909 \AA\ and C\,\textsc{iv} 1549 \AA , with its associated broad absorption feature bluewards of C\,\textsc{iv}. 
Conversely, the $r$ band covers mainly the continuum and the Fe\,\textsc{ii} and Fe\,\textsc{iii} pseudo-continua. The impact of all the aforementioned spectral features will be discussed further in Sec. 5.1.

%--------------------------------------------------------------------

\section{Variability features and methods}

\subsection{Excess variance $\sigma_{rms}^2$}
We aim to establish the difference between variability amplitudes of BALs and non-BALs. As a first step, we compare their normalized excess variances (\citealt{Nandra1997}; \citealt{Turner1999}; \citealt{Allevato2013}; \citealt{Sanchez2017}), which capture variability across all timescales sampled by the light curves and provide a measure of the intrinsic variability amplitude after accounting for measurement noise. Following the definition

\begin{equation}
    \sigma^2_{rms} = \frac{1}{N_{obs}\bar{x}^2}\sum^{N_{obs}}_{i=1}[(x_i-\bar{x})^2-\sigma^2_{err,i}],
\end{equation}

where $x_i$ and $\sigma_{err,i}$ are the count rate and its associated measurement uncertainty at the $i$-th epoch of the light curve, $\bar{x}$ is the mean count rate, and $N_{\mathrm{obs}}$ is the number of observations used to estimate $\sigma^2_{rms}$. 

The uncertainty on $\sigma^2_{rms}$ due to Poisson noise is given by

\begin{equation}
    err(\sigma^2_{rms})=\frac{S_D}{\bar{x}^2N^{1/2}_{obs}},
\end{equation}

\begin{equation}
    S^2_D=\frac{1}{N_{obs}}\sum^{N_{obs}}_{i=1}\{[(x_i-\bar{x})^2-\sigma^2_{err,i}]-\sigma^2_{rms}\bar{x}^2\}^2.
\end{equation}

To calculate the excess variance, we first converted the observed ZTF magnitudes into linear flux units. For each epoch, the flux was calculated from the observed magnitude, $m$, using the standard relation $F = 3631\times 10^{-0.4 m}$Jy, with the scaling factor derived from the AB magnitude system zero point.
The proportionality constant and units go away when normalizing the excess variance by the median flux, squared. We then propagated the magnitude errors ($\delta_m$) to obtain the corresponding flux uncertainties via $\delta_F = \frac{\ln(10)}{2.5} F \delta_m$.  For the calculations done with the DRW we used the original magnitudes, as this code does not normalize by mean flux.

\subsection{DRW modelling}

Optical variability in AGNs is commonly modelled as a DRW process \citep{Kelly2009}. The DRW model depends on two free parameters: the driving amplitude $\sigma_{\mathrm{DRW}}$, whose square corresponds to the variance of the light curve in the stationary limit, and the damping timescale $\tau_{\mathrm{DRW}}$, which marks the characteristic variability timescale.
In this work, we used the \texttt{turbo-fats} Python package 
\footnote[1]{Code repository: \url{https://github.com/alercebroker/turbo-fats}}
to extract these metrics. This software implements the DRW 
model through Gaussian process (GP) regression following \citet{Graham2017}. The GP uses an Ornstein-Uhlenbeck kernel, which defines the covariance between two observations of the light curve separated by a time lag, $\Delta t$, as

\begin{equation}
    k(\Delta t) = \sigma^2_{\mathrm{DRW}} \exp\left(-\frac{|\Delta t|}{\tau_{\mathrm{DRW}}}\right).
\end{equation}

In this framework, $\sigma^2_{\mathrm{DRW}}$ is the variance of the light 
curve, corresponding to the total variability amplitude in the stationary 
limit ($\Delta t \gg \tau_{\mathrm{DRW}}$), where the covariance evaluated 
at zero lag gives

\begin{equation}
    k(0) = \sigma^2_{\mathrm{DRW}} = \mathrm{Var}(m).
\end{equation}

By measuring $\sigma_{\mathrm{DRW}}$ and $\tau_{\mathrm{DRW}}$ across the BAL and non-BAL populations, we can compare the long-term variance and characteristic timescales of both populations. Given that the power spectral break timescale is known to correlate with \mbh\ and \redd\ \citep[e.g.][]{Burke2021, Arevalo2024}, it allows us to test for differences in the physical  properties of our sub-samples using a method entirely distinct from the single-epoch mass measurements previously described for this sample.

\subsection{Bootstrap procedure}
To estimate the median values of our metrics and their associated uncertainties, we implemented a bootstrap procedure using 1,000 resamplings. This method simulates repeatedly sampling the underlying population by generating multiple 
resamples and computing the median for each iteration. The median of the resulting bootstrap distribution is adopted as the final central value, and its standard deviation defines the statistical uncertainty. For the excess variance, given that 
we have individual measurement errors provided by Equation 2, we complemented the 
bootstrapping with a Monte Carlo (MC) procedure. Over the 1,000 iterations, each 
individual $\sigma_{\mathrm{rms}}^2$ value is randomly perturbed according to its 
error, ensuring that both sample variance and measurement uncertainties are fully 
propagated for this metric.

\section{Results}
\label{sec:results}
\begin{figure}[!h]
    \centering
    \includegraphics[width=0.9\linewidth]{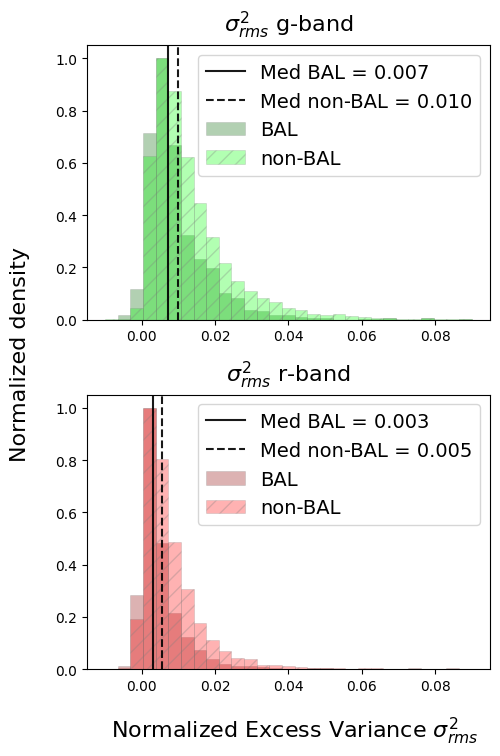}
    \caption{Normalized excess variance ($\sigma_{\mathrm{rms}}^{2}$) distributions for the BAL and non-BAL quasar samples, shown for the $g$ (top) and $r$ (bottom) bands. The median variance of each population is indicated by vertical lines. BAL quasars show consistently lower $\sigma_{\mathrm{rms}}^{2}$ values compared to their non-BAL counterparts. }
    \label{fig:xs}
\end{figure}

To assess the overall variability amplitudes of both quasar populations, we first compare their normalized excess variance distributions (Fig.~\ref{fig:xs}). While the two samples largely overlap, the median values measured using the bootstrap and MC procedure show a clear offset, indicating lower variability among BAL quasars. In the $g$ band, we find population medians of $(70.8 \pm 2.1 )\times 10^{-4}$ and $(99.5 \pm 1.3) \times 10^{-4}$ for BAL and non-BAL quasars, respectively, while the corresponding values in the $r$ band are $(32.2 \pm 1.3)\times 10^{-4}$ and $(55.1 \pm 0.9)\times 10^{-4}$. These results indicate that non-BAL quasars have systematically higher variability than their BAL counterparts across both filters. Additionally, both populations show increased excess variance in the $g$ band relative to the $r$ band.

To fully characterize the physical overlap between the samples, we also measured their intrinsic scatter using the interquartile range (IQR). The $g$-band IQR spans 0.00867 for BALs and 0.01185 for non-BALs, while the $r$-band IQR is 0.00522 and 0.00809, respectively. Because of this wide dispersion, the median variability of the BAL population falls entirely within the non-BAL IQR. However, this large intrinsic scatter among individual quasars does not diminish the statistical significance of the population-level offset. A two-sample Kolmogorov–Smirnov (KS) test confirms that the visual separation in variability is fundamentally distinct. The KS test measures the maximum separation ($D$) between the cumulative distributions of both samples, yielding $D = 0.185$ in the $g$ band and $D = 0.211$ in the $r$ band.  Given our large sample size, a separation of this magnitude is not a marginal or modest fluctuation amplified by statistics; rather, it represents a highly robust and physically meaningful distinction between the two populations. Supported by near-zero $p$-values ($\sim10^{-27}$ and $\sim10^{-35}$, respectively), these results demonstrate that despite the individual-source overlap, BAL quasars are less variable as a population.

The variability amplitude and the break timescale are known to be a function of \mbh\ and \redd\ \citep[e.g.][]{Arevalo2023}, and the distributions of these properties are not identical for the BAL and non-BAL samples, as shown in Fig. \ref{fig:properties}. Therefore, we must explore whether the differences in variance persist when comparing objects with identical properties. To this end, we split the sample into a 3 $\times$ 3 grid of bins across the overlapping parameter space.
 Specifically, we defined three bins in black hole mass spanning from 8.8 to 9.7 in $\log\, $\mbh, and three bins in Eddington ratio spanning from -1.2 to -0.3 in $\log\,$\redd. A constant bin width of 0.3 dex was used for both parameters. The bin limits and the number of light curves included in each bin are summarized 
in Table \ref{tab:lc_per_bin}. The individual $\sigma_{\mathrm{DRW}}$ and $\tau_{\mathrm{DRW}}$ metrics measured for the objects in each range are compiled within these bins for the subsequent analysis.
To determine the representative values and errors for all parameters within each 
bin we 
applied the bootstrap procedure mentioned above. To ensure robust statistics during 
the bootstrap resampling, we originally imposed a minimum threshold of ten 
quasars per bin.

\begin{figure*}[!h]
    \centering    
    \includegraphics[width=0.9\linewidth]{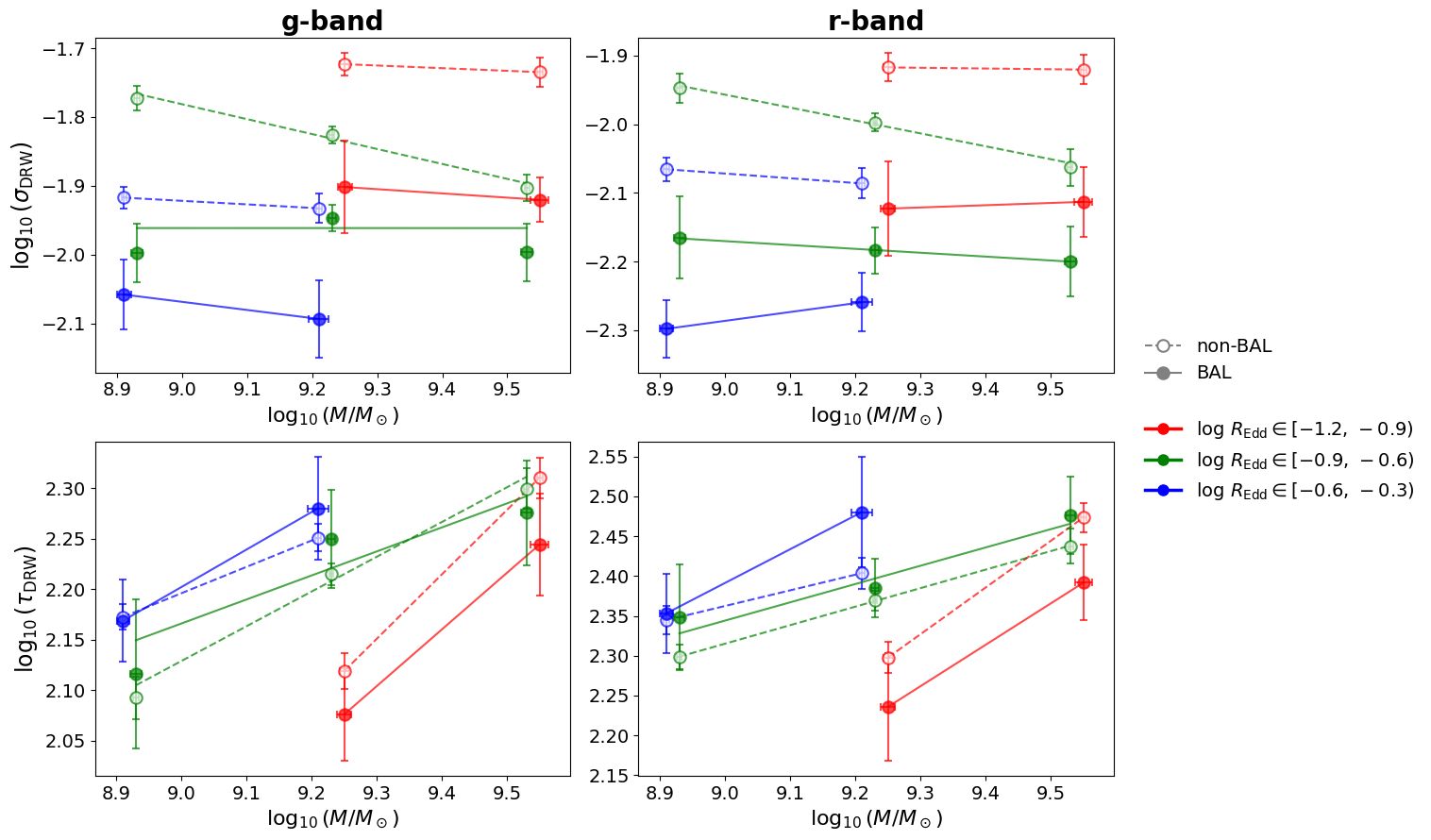}
    \caption{DRW parameters from the $g$- and $r$-band light curves as a function 
of \mbh\ for three \redd\ intervals. Each colour corresponds to a different 
\redd\ bin. Solid markers represent median BAL values per bin, and open 
markers represent non-BAL median values. Within each \redd\ range, the points 
have been slightly offset horizontally by 0.02 for visualization purposes.}
    \label{fig:bins_mh}
\end{figure*}

Within this grid, we plotted the median values for each \mbh\ bin within every 
\redd\ bin (see Fig. \ref{fig:bins_mh}). Because we require a minimum of ten light 
curves to populate a bin, some bins lack a valid counterpart; consequently, the 
plot only includes paired bins that share the same mass range across different 
\redd\ ranges, allowing us to compute the ratios shown later in our analysis. Due 
to this paired selection requirement for the parity comparison, the point with the 
fewest light curves among the retained bins actually contains 71 
light curves.

Each row of panels in Fig. \ref{fig:bins_mh} represents a DRW 
metric as a function of \mbh\ for the three \redd\ intervals described in Table 
\ref{tab:lc_per_bin}. Each column corresponds to the data for a given band, with 
the data points representing the median $\sigma_{\mathrm{DRW}}$ and $\tau_{\mathrm{DRW}}$ 
values of the BAL and non-BAL samples in the corresponding \mbh\ and \redd\ bins.
A linear fit is plotted in each case to illustrate the overall trend of 
he metrics with \mbh\ and to compare the two populations directly. 
These fits were computed using orthogonal distance regression (ODR) weighted by the 
uncertainties in both axes; therefore, points with higher precision dominate the 
fit, occasionally offsetting the line from the geometric centre of the data points.
As can be seen in the top row, the long term variance ($\sigma_{\mathrm{DRW}}$) is largely independent of \mbh, but increases with decreasing \redd, as expected. This is seen for both the BAL and non-BAL samples. However, the values of $\sigma_{\mathrm{DRW}}$ are consistently lower for the BAL sample when compared to non-BALs of the same \mbh\ and \redd\ (i.e., the solid lines are always below the dashed lines of the same colour), indicating reduced variability at fixed \mbh\ and \redd.

For the bottom row panels, the characteristic timescale of the variations increases as a function of \mbh\ and also with increasing \redd, as is expected according to \citet{Burke2021} and \citet{Arevalo2024}. However, the values of $\tau_{\mathrm{DRW}}$ do not show a significant difference between the BAL and non-BAL bins, with overlapping error bars in all bins.
Additionally, their small differences are not constant, with the BALs showing larger or smaller values of $\tau_{\mathrm{DRW}}$ depending on the specific \redd\ bin. We note that the interpretation of the actual values of $\tau_{\mathrm{DRW}}$ is complicated by possible biases, we discuss these below in Sec. \ref{sec:biased_mass}. 

For each band, the ratio between the median variance of BAL and non-BAL quasars was computed for all seven bins where both samples met the selection criteria. 
The mean of these seven ratios provides an estimate of the average relative variability amplitude between the two populations. 

As summarized in Table \ref{tab:drw_ratios}, in the $g$ band, BAL quasars display only $\sim 70\%$ of the variance of similar non-BAL quasars, and in the $r$-band BAL quasars display only $\sim 64\%$ of the variance seen in non-BAL quasars. In contrast to this amplitude  suppression, their characteristic timescales are essentially the same. The  $\tau_{\mathrm{BAL}} / \tau_{\mathrm{non\text{-}BAL}}$ ratio is $0.981 \pm 0.081$  for the $g$ band and $1.017 \pm 0.128$ for the $r$ band. With both measurements falling within $1\sigma$ of unity. In the appendix \ref{sec:appendix} we state the median luminosity and redshift in each grid cell, showing that these values are very similar for both samples.

\begin{table}
\centering
\small
\caption{Number of light curves per sample and bin}
\resizebox{\columnwidth}{!}{%
\begin{tabular}{lcccc}
\hline\hline
 & \multicolumn{4}{c}{\hspace{1.7cm}Range in $\log\, R_{\mathrm{Edd}}$} \\
\cline{3-5}
 & & $-1.2$ to $-0.9$ & $-0.9$ to $-0.6$ & $-0.6$ to $-0.3$ \\
\hline
\multicolumn{5}{l}{non-BAL}\\
\multirow{3}{*}{\parbox{2cm}{\raggedright Range in log \mbh }} 
 & 8.8--9.1 & 22 & 571 & 887 \\[-2pt]
 & 9.1--9.4 & 481 & 1832 & 547 \\[-2pt]
 & 9.4--9.7 & 706 & 613 & 24 \\
 \\
\multicolumn{5}{l}{BAL}\\
\multirow{3}{*}{\parbox{2cm}{\raggedright Range in log \mbh}} 
 & 8.8--9.1 & 0 & 64 & 123 \\[-2pt]
 & 9.1--9.4 & 71 & 259 & 71 \\[-2pt]
 & 9.4--9.7 & 135 & 115 & 0 \\
\hline
\end{tabular}%
}
\begin{tablenotes}
\item {\bf Note.} Number of light curves per \mbh–\redd bin for the non-BAL and BAL samples, shown in the top and bottom grids, respectively. Bin limits are given along the rows and columns.
\end{tablenotes}
\label{tab:lc_per_bin}
\end{table}

%%%%

\begin{table}[ht]
\centering
\caption{Mean DRW parameter ratios between BAL and non-BAL samples}

\begin{tabular}{lcc}
\hline\hline
 & Metric & $\langle \mathrm{BAL} / \mathrm{non}\text{-}\mathrm{BAL} \rangle$ \\
\hline
\multirow{2}{*}{\parbox{1cm}{\centering $g$ band}} 
 & $\sigma_{\mathrm{DRW}}$ & $0.697 \pm 0.065$ \\[-2pt]
 & $\tau_{\mathrm{DRW}}$  & $0.981 \pm 0.081$ \\
 \\[-4pt]
\multirow{2}{*}{\parbox{1cm}{\centering $r$ band}} 
 & $\sigma_{\mathrm{DRW}}$ & $0.642 \pm 0.043$ \\[-2pt]
 & $\tau_{\mathrm{DRW}}$  & $1.017 \pm 0.128$ \\
\hline
\end{tabular}
\begin{tablenotes}
\item {\bf Note.} Comparison of DRW parameters for the $g$ and $r$ bands. The reported values were derived by first calculating the BAL/non-BAL ratio within each of the seven comparable ($M_{\mathrm{BH}}$, $R_{\mathrm{Edd}}$) bin pairs, and then averaging those seven individual values.
\end{tablenotes}
\label{tab:drw_ratios}
\end{table}

%%%%%

\section{Discussion}
\label{sec:discussion}
The results presented in the previous section reveal a consistent and significant difference in the variability amplitudes of BAL and non-BAL quasars. In every comparison, BAL quasars show systematically lower variance values than their non-BAL counterparts. This trend is evident both in the overall $\sigma_{rms}^2$  distributions for the $g$ and $r$ bands (Fig.~\ref{fig:xs}) and in the binned analysis as a function of \mbh\ and \redd\ (Fig.~\ref{fig:bins_mh} ).
This behaviour persists across all \redd\ intervals, indicating that BAL  quasars having systematically less variability is a robust characteristic.
The mean ratios of BAL to non-BAL median variances, summarized in Table~\ref{tab:drw_ratios}, indicate that BAL quasars have, on average, only $\sim$70\% of the variance of the non-BAL quasars, in the $g$ band, and $\sim$64\% of the variance in the $r$ band.

It is worth noting that these differences cannot be attributed to systematic biases associated with redshift or rest-frame wavelength coverage. As shown in Fig.~\ref{fig:properties}, the redshift distributions of the BAL and non-BAL samples are closely matched, implying that both populations are sampled over similar rest-frame wavelength ranges and that the observed timescales correspond to comparable intrinsic timescales.
Therefore, the observed discrepancy in variability amplitude must arise from intrinsic physical differences between the two populations rather than from observational or selection effects.

\subsection{Effect of the absorption and emission lines}

As shown in Fig. \ref{fig:spec_bal}, the $g$- and $r$-band light curves trace the variations of different parts of the spectrum. Broad emission lines and broad absorption features in BALs are concentrated in the $g$ band at typical sample redshifts, whereas the $r$ band primarily captures the continuum. The presence of the variability suppression in both optical bands already suggests that the effect is not solely driven by the behaviour of BAL troughs or emission lines. However, because the spectral features sampled by the photometric bands vary with redshift, it is necessary to examine whether line variability could contribute to the observed trends.

\subsubsection{Assessing the contribution of emission and absorption features }

Large-scale statistical studies of BAL spectral variability \citep[e.g.][]{Wang2015, He2017} show that more than $\sim$70\% of BAL troughs exhibit a negative response to continuum variations, in the sense that the absorption strength decreases when the ionizing continuum brightens. This behaviour is naturally expected if the BAL gas is predominantly in a high-ionization state, where fluctuations in the ionizing flux efficiently ionize species such as C\,\textsc{iv} into higher stages, weakening the observed absorption. Recent work by \citet{He2025} also highlights an asymmetric response in this photoionized outflowing gas due to recombination timescales and ionization rates.

If this photoionization-driven absorption variability were the dominant driver of flux variations, BAL quasars would display larger variability amplitudes than non-BAL quasars in bands covering the troughs. This is because as the continuum flux increases, the absorption strength decreases, leading to an even larger increase in the flux within a photometric band, while the opposite would happen as the flux decreases. However, for the general population, this expectation is contradicted by our results: BAL quasars are systematically less variable than non-BALs, indicating that photoionization effects cannot explain the overall suppression of variability. Instead, it plays a secondary, localized role: The rapid response of high-ionization BAL gas allows changes in the absorption strength to be imprinted into the observed flux, increasing the total variance specifically within bands covering the absorption troughs. This reduces the variability contrast between the two populations in the $g$ band compared to the $r$ band, yielding the higher values of $\langle \sigma_{\mathrm{BAL}} / \sigma_{\mathrm{non-BAL}} \rangle$ summarized in Table~\ref{tab:drw_ratios}.

On the other hand, the $r$ band covers the continuum and the Fe\,\textsc{ii} and Fe\,\textsc{iii} pseudo-continua. Since these iron lines are characterized by their relative stability, their contribution may suppress the relative amplitude of variability. In the UV, narrow-band light curves covering the Mg\,\textsc{ii} and Fe\,\textsc{ii} complex (2700–2900\,\AA\ rest-frame) can exhibit about half the amplitude of variability as neighbouring narrow bands, as shown for the Quasar CTS C30.10 by \citet{Prince2022}. Similarly, in the optical regime, the Fe\,\textsc{ii} complex shows variations roughly 10\% lower than those of H$\beta$ and about half the amplitude of the continuum variations, judging from the light curves presented by \citet{Hu2015} for ten AGN.

To evaluate the specific impact of this dilution within our own dataset, we computed the flux contribution of the iron emission complex. Based on the \texttt{FEII\_UV\_EW} parameter from \citet{Lyke2020}, the mean Fe\,\textsc{ii} EW in our sample is $\sim$35\,\AA\ for both BAL and non-BAL QSOs. 
Adopting a mean EW of $\sim$13\,\AA\ for the adjacent Fe\,\textsc{iii} emission \citep[derived from a lower-redshift sample at $z < 0.8$;][Martínez-Collipal et al., in prep.], the total contribution of the iron pseudo-continua to the $r$ band at $z=1.8$ ($\sim$625\,\AA\ rest-frame effective width) is only $\sim$7.6\%. To estimate the maximum effect that a constant Fe emission could have on the difference between both populations, we assumed an extreme scenario: the Fe complex remains completely stable in BALs (causing maximum dilution), but is as variable as the continuum in non-BALs (causing zero dilution). In this case, the constant Fe pseudo-continuum would dilute the continuum variance in BALs to
$\left(\frac{\delta F}{F}\right)^2 = \frac{1}{(1-0.076)^2} \left(\frac{\delta C}{C}\right)^2 \approx 85\%$
of the intrinsic value. Therefore, even this maximum possible difference from a constant Fe pseudo-continuum cannot explain the entire drop in variance seen between the BAL and non-BAL samples mentioned in Sec. \ref{sec:results} It can contribute a $\sim$15\% drop, but it is insufficient to account for the reported differences.

\subsubsection{Isolating the effect of spectral features via redshift and luminosity dependencies}

\begin{figure*}[!h]
    \centering    
    \includegraphics[width=0.9\linewidth]{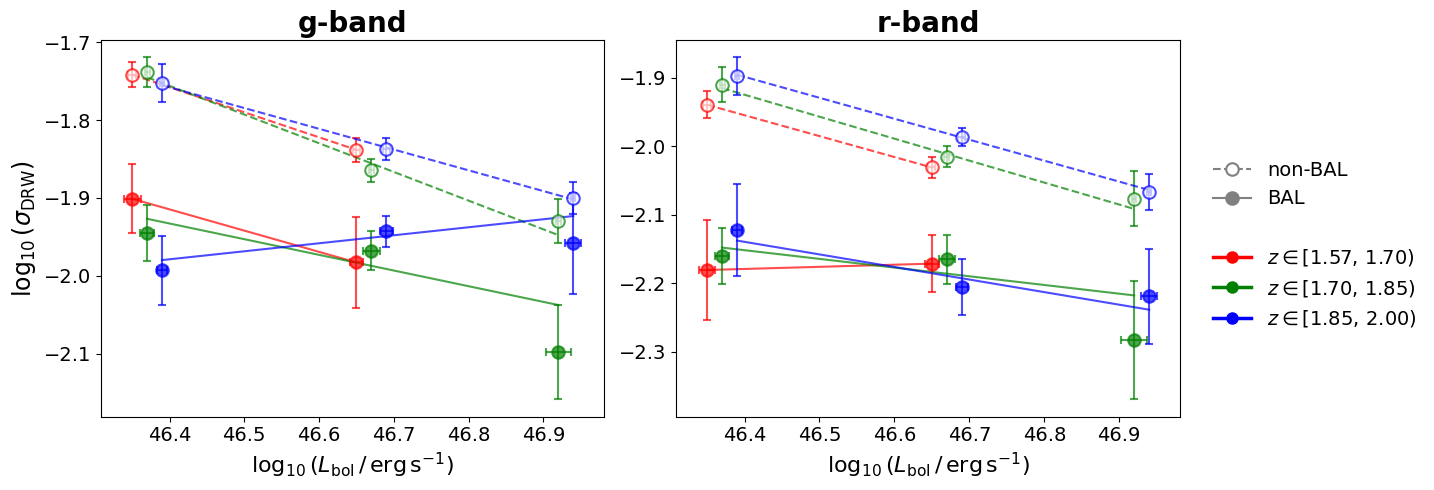}
    \caption{
    Median $\sigma_{\mathrm{DRW}}$ values per bin from the $g$- and $r$-band light curves as a function of $L_{\mathrm{bol}}$ for three redshift intervals. As in Fig.~\ref{fig:bins_mh}, each colour corresponds to a different redshift bin. Solid markers represent BAL median values, and open markers represent non-BAL median values.}
    \label{fig:bins_zl}
\end{figure*}

To explicitly test whether these shifting spectral features dominate our observations or whether the suppression is truly intrinsic to the continuum, we analysed how $\sigma_{\mathrm{DRW}}$ behaves as a function of both redshift and luminosity. This approach serves a dual purpose. First, it allows us to verify if our sample follows the well-established global trend wherein quasar variability decreases with increasing luminosity. Second, by slicing the sample into redshift intervals, we can track spectral features as they pass through our photometric filters. The redshift bins limits are 1.57--1.7, 1.7--1.85, and 1.85--2.0. Therefore, the $r$ band covers the C\,\textsc{iii} emission line only in the highest redshift bin, and there are no strong emission lines in this band in the other two redshift bins. On the other hand, the $g$ band contains the C\,\textsc{iv} absorption trough wavelengths in the highest redshift bin, and in about half the objects in the intermediate redshift bin, while all redshift bins contain either C\,\textsc{iii} or C\,{\sc iv} emission lines, or both.

\begin{figure}[!h]
    \centering    \includegraphics[width=0.9\linewidth]{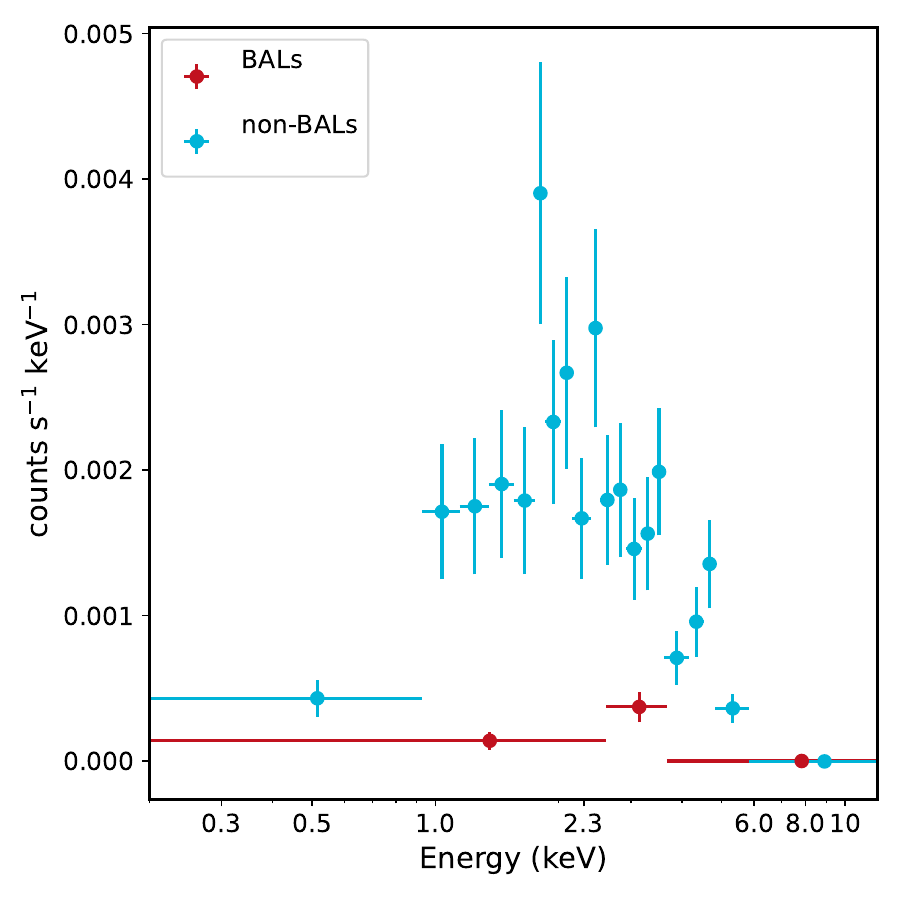}
    \caption{Stacked X-ray spectra of the BAL sample and a random non-BAL subsample, of 310 objects each, extracted from eROSITA-DE DR1 exposures. The spectra are binned to have a minimum of 20 source counts each. }
    \label{fig:stack_xray}
\end{figure}

The results of this breakdown are presented in Fig. \ref{fig:bins_zl} (which displays the $\sigma_{\mathrm{DRW}}$ median values per bin using the same format as Fig. \ref{fig:bins_mh}, but this time split into a grid of luminosity and redshift). In the $r$ band, $\sigma_{\mathrm{DRW}}$ remains consistently lower for BAL quasars across all luminosity bins, and all redshift bins, regardless of whether C\,\textsc{iii} falls inside the wavelength range or not. This persistent offset confirms that shifting emission lines do not heavily distort our $r$-band results, validating our previous calculation: when looking at pure continuum, the suppression of variability in BALs is a robust feature.

In contrast, the $g$ band reveals the localized influence of the C\,\textsc{iv} absorption line responsivity discussed above. For all luminosity bins in the low and intermediate redshifts we see the suppression of variability in BALs. Only in the highest redshift range ($1.85 < z < 2.00$) and at the highest luminosity bin, the characteristic suppression of variability in BALs mostly disappears, with both populations exhibiting comparable variance levels. In this window, the pronounced continuum suppression is masked by the additional flux fluctuations induced by the trough's photoionization dynamics.

Overall, our results show that neither BAL absorption variability nor emission-line contamination can account for the lower $\sigma_{\mathrm{DRW}}$ values observed in BAL quasars. While these spectral features can influence the measured variability in particular photometric bands, the suppression persists in the continuum-dominated $r$ band, indicating that the effect is intrinsic to the continuum variability. In the following section, we explore several observational properties of BAL quasars that could potentially drive the observed suppression in variability.

\subsection{Potential Drivers for the Suppressed Variability in BAL QSOs}

\subsubsection{Biased \mbh\ estimates}
\label{sec:biased_mass}
We observe similar trends of variance versus \mbh\ and \redd\ in the BAL and non-BAL samples, albeit with lower normalizations for the BAL sample. Therefore one possibility to explain the difference in variance is that the \mbh\ of the BALs are overestimated, and therefore the \redd\ are underestimated. We note that for this work the \mbh\ of all the objects were estimated using the single epoch method for the Mg\,\textsc{ii} 2798 \AA\ line. Although this transition is much less affected by broad absorption features and outflows than C\,\textsc{iv}, observational and modelling studies have shown that Mg\,\textsc{ii} emission can be significantly influenced by non-virial components associated with nuclear outflows, affecting the line width and shift, thereby biasing virial mass estimates \citep[e.g.][]{Marziani2013,Liu2019,Yi2020}. A systematic overestimation of the \mbh\ in the BAL but not in the non-BALs, and consistent underestimation of the \redd, would produce a mismatch in the variances in the same direction that is observed. This is because there is a known, strong anti-correlation between \redd\ and optical/UV variance \citep[e.g.][]{MacLeod2010,Zuo2012,Rakshit2017,Arevalo2023}. Therefore, if the \redd\ of only the BALs, are indeed much higher than the current estimates, then their lower variability would fit the behaviour of the rest of the quasar population. 

\begin{figure*}
    \centering
    \includegraphics[width=0.9\linewidth]{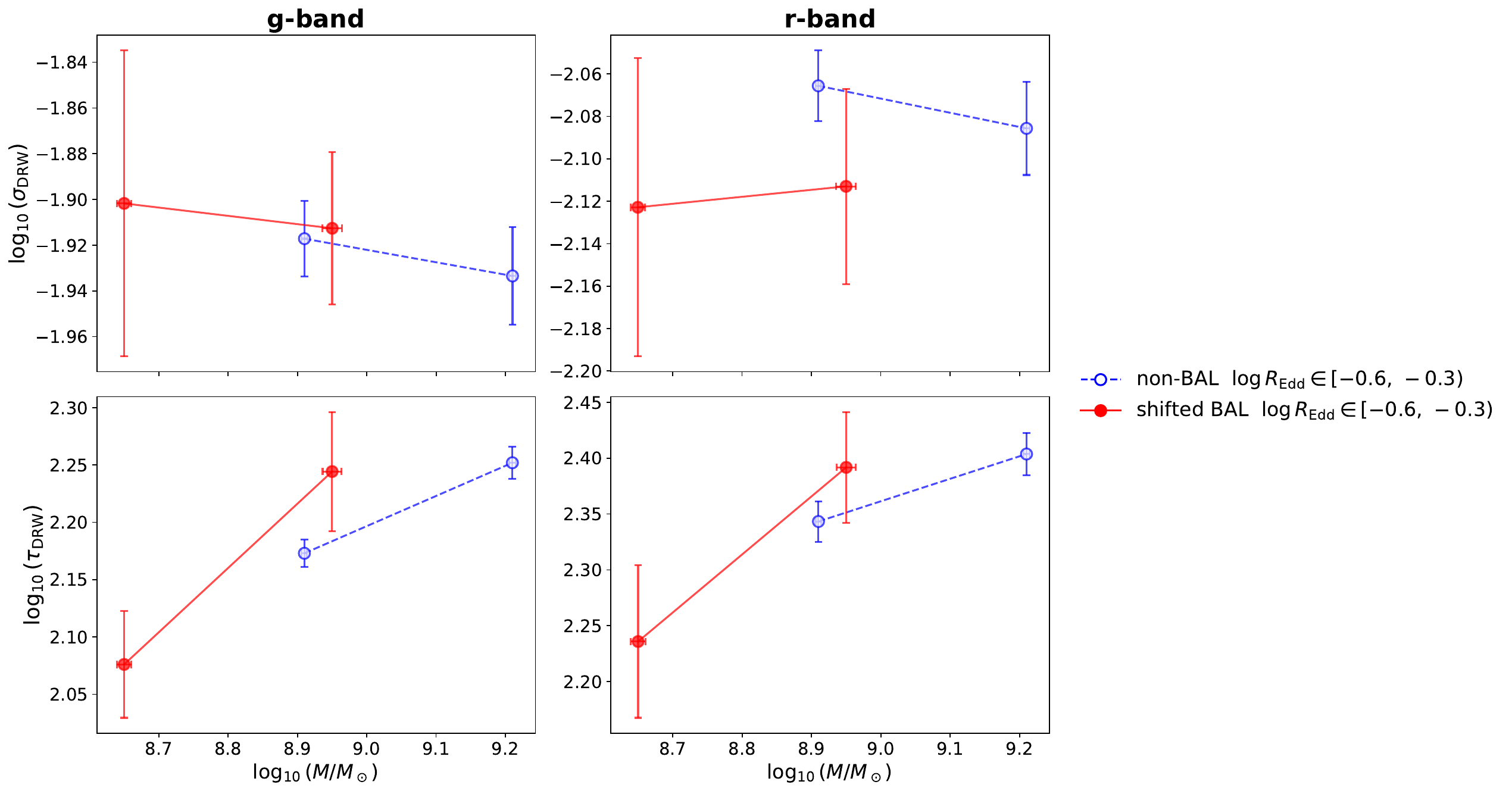}
    \caption{Test of the \mbh-bias hypothesis on DRW variability parameters: to evaluate whether the reduced variability in BAL quasars could be an artefact of a systematic \mbh\ overestimation in this population, a shift of $\Delta \log$ \mbh $= -0.6$ (and corresponding $\Delta \log$ \redd $= +0.6$) is applied to the BAL sample. For clarity, only the parameter grid points that are directly matched in both \mbh\ and \redd\ after the shift are displayed. Solid red markers represent the BAL sample after shifting their \mbh\ and \redd , while open blue markers represent the non-BAL sample in its original form.}
    \label{fig:bins_DRW_corrected}
\end{figure*}

We performed the following simple experiment to estimate the magnitude of the \mbh\ overestimation that would be needed to explain the difference in variance. Shifting all the \mbh\ of the BAL sample down by $\Delta \log $\mbh$=-0.6$ moves all the BAL markers in Fig. \ref{fig:bins_mh} two places to the left. At the same time, since the luminosity does not change, the \redd\ estimates move up by $\Delta \log $\redd$=+0.6$, so that the intervals in \redd\ shift by two, i.e. the lowest interval in \redd\ in the BAL sample should now be compared to the highest interval in the non-BAL sample. These shifts produce a better match between the BAL and non-BAL variances, as shown in Fig. \ref{fig:bins_DRW_corrected}. These differences in the \mbh\ and \redd\ distributions of the BAL and non-BAL samples do require a large, i.e. $\sim 0.6$ dex, equivalent to a factor of 4, systematic overestimation of the \mbh\ in BALs compared to the non-BALs.

A systematically higher \redd\ for the BAL population is physically consistent with current theoretical frameworks. For instance, according to the failed radiatively accelerated dusty outflow (FRADO) models \citep{Naddaf2023}, the geometry of the wind strongly depends on the accretion rate. As the Eddington ratio increases, the solid angle from which these outflows are launched expands. Consequently, high-\redd\ sources present a larger covering factor for winds, significantly increasing the probability that an observer's line of sight will intercept the outflow and therefore detect it as BALs.

A biased \mbh\ scenario  would also affect the comparison of the characteristic timescales, $\tau_{\mathrm{DRW}}$, which, as shown before, were already very consistent for BALs and non-BALs of the same \mbh\ and \redd . Because the characteristic timescale of quasar optical variability scales positively with black hole mass \citep[e.g.][]{Burke2021,Arevalo2024}, any true systematic discrepancy in $M_{\mathrm{BH}}$ of this magnitude ($\sim$0.6~dex) could manifest as a noticeable separation in the $\tau_{\mathrm{DRW}}$ values of the two populations. Specifically, comparable bins with the same \mbh\ and \redd\ should yield the same $\tau_{\mathrm{DRW}}$ values; conversely, if the \mbh\ were systematically overestimated, the BAL bins 
would show a larger $\tau_{\mathrm{DRW}}$ than the non-BAL bins.

However, as shown in Fig. \ref{fig:bins_DRW_corrected}, a comparison between the measured $\tau_{\mathrm{DRW}}$ values of matched, corrected \mbh\ and \redd\ bins shows that no statistically significant differences are exhibited between the BAL and non-BAL samples.
Nevertheless, these $\tau_{\mathrm{DRW}}$ comparisons must be interpreted with caution due to the inherent limitations of DRW modelling over finite baselines. As shown by \citet{Kozlowski2017}, robust recovery of the characteristic timescale generally requires an observation baseline (T) significantly longer than the timescale itself ($T / \tau_{\mathrm{DRW}} \gtrsim 10$). For our sample, the $\sim 6$-year ZTF observed baseline corresponds to a median rest-frame baseline-to-$\tau_{\mathrm{DRW}}$ ratio of $T / \tau_{\mathrm{DRW}} \approx 4.5$ across our redshift range ($1.57 \le z \le 2.00$). In this regime, individual $\tau_{\mathrm{DRW}}$ estimates carry higher uncertainty, meaning the overlapping timescales between BALs and non-BALs could be an artefact of our limited baseline length and sample size rather than proof of identical physical properties. While this lack of separation challenges the \mbh-bias argument, a subtle timescale shift cannot be fully ruled out until longer baselines and larger sample sizes reduce these combined observational and statistical uncertainties.

\subsubsection{Nuclear obscuration} 

Beyond accretion physics, the consistently lower variability in our BAL sample -- which persists even at fixed \mbh~and \redd\ -- may be driven by geometric orientation and wind reprocessing rather intrinsic accretion differences. This aligns with \citet{Welling2014} and radio spectral index studies \citep[e.g.][]{Bruni2012, DiPompeo2012}, which suggest BALs are often viewed at higher inclinations (steeper $\alpha_r$).
This scenario is analogous to the obscuration seen in Type 2 AGNs, which exhibit significantly lower optical variability than Type 1s because the dusty torus conceals the central engine \citep{Kovacevic2025} and the main source of optical flux in a Type 2 AGN is the stable starlight of the host galaxy. A similar process might occur in BAL quasars, which are characteristically redder than their non-BAL counterparts \citep{Reichard2003, Gibson2009, Petley2022}; however, while the torus is the primary obscurer in Type 2 sources, the reddening and suppressed variability in BALs would then originate from dust embedded within the outflows. However, since the variability amplitudes we measure are relative to the total flux, and this UV flux is not dominated by the host galaxy, the obscuration would have to act preferentially on the portion of the accretion disc that is intrinsically more variable. 
Furthermore, the BAL wind itself likely absorbs variable EUV/X-ray photons and re-emits them as broad lines or thermal emission; this reprocessing would effectively 'dampen' the sharp continuum fluctuations seen in the optical $g$ and $r$ bands.

\subsubsection{Reduced X-ray reprocessing}

Continuum reverberation in AGN measures the delayed response of the accretion disc emission in different wavelengths, emitted predominantly at different radii in the disc. These delays are interpreted as the reprocessing of variable X-ray irradiation from a compact corona \citep[e.g.][]{McHardy2014,Cackett2022,Papoutsis2024}. In this framework, a fraction of the coronal X-ray emission is intercepted by the disc and reprocessed into UV and optical radiation, imprinting correlated variability across these wavelengths. The efficiency of this mechanism to generate optical/UV variability therefore depends directly on the incident X-ray flux. Since BAL quasars are known to be systematically X-ray weak compared to non-BAL quasars at similar optical/UV luminosities \citep[e.g.][]{Green2001,Gallagher2006,Stalin2010,Liu2018}, the contribution of X-ray reprocessing to their UV variability can be expected to be reduced, if the disk indeed receives less X-ray flux. The X-ray weakness of BALs can therefore provide a natural explanation for the lower continuum variability amplitudes observed in BAL quasars.

It is important to clarify the cause of the observed X-ray weakness however, which can be produced by strong absorption, or can instead be intrinsic, to interpret the significance of the variability results. Studies of the X-ray spectrum of BALs have concluded that although they are observed to be X-ray weak compared to their UV luminosity, once the effect of absorption has been taken into account, the X-ray luminosities appear consistent with those of non-BAL quasars \citep{Green2001,Gallagher2006}. In contrast, other authors have found evidence for intrinsic X-ray weakness in BALs. For example,  \citet{Liu2018} estimate that 6-23\% of BALs are intrinsically X-ray weak from deep Chandra observations of a sample of BAL quasars, and \citet{Luo2014} conclude that at least half their sample of BALs observed with the hard X-ray telescope NuSTAR are indeed unobscured, weak X-ray emitters. Finally, \citet{Yang2021} also conclude that even after accounting for absorption, the X-ray flux of a high-z, lensed BAL is several times below the level expected for normal quasars. In the latter case, the weaker UV variability in BALs is expected given the lower X-ray flux available to produce variability of the accretion disc emission through reprocessing. In the obscuration scenario, the obscuring material should be positioned in a way that can block the view between the X-ray corona to the accretion disc as well as to the observer.

In either case, we assessed the observed X-ray properties of our samples, by searching for counterparts in the eROSITA-DE DR1 catalogue of X-ray sources, which covers half the sky, at Galactic longitudes of $\ell > 180^\circ$ \citep{Merloni2024}. Our final sample contains 1937 non-BAL and 310 BAL objects in this region of the sky. Performing a positional cross-match with the eROSITA-DE catalogue using a matching radius of 10~arcsec, results in 404 detections for the non-BAL sample and only five detections for the BAL sample. This corresponds to X-ray detection fractions of $21\%$ for non-BAL quasars and $1.6\%$ for BAL quasars, respectively. The contrast between the detection rates is fully consistent with the known X-ray weakness of BAL QSOs. If the reprocessing mechanism is less efficient due to a lack of incident X-ray flux, the BAL continuum would appear significantly more stable, consistent with our reported results.

In addition, we estimated the difference in X-ray fluxes by stacking  observations from the eROSITA DR1 of both samples. We extracted source and background counts for 310 BALs in the eROSITA-DE footprint and for a randomly selected control subsample of 310 non-BAL sources, using the task \texttt{SRCTOOL} contained in the eROSITA reduction package \texttt{eSASS}. The redshift of each target was used to select photons in a common restframe energy range before stacking the spectra with the \texttt{XStack} software 
package \citep{Chen2025}. We estimated the flux of each sample by fitting an absorbed power law model, with a fixed photon index $\Gamma=1.9$ and tied absorbing column density between both spectra, using Cash statistics to select the best fit. The resulting power law normalizations were $N=9.6^{+5.8}_{-5.6}\times 10^{-7}$ for the BAL sample and $N=77^{+11}_{-6}\times 10^{-7}$ for the non-BALs. This difference by a factor of 8 in the normalization of the stacked X-ray spectra shows that in our samples too the BALs are significantly X-ray weaker (see Fig.~\ref{fig:stack_xray}). For completeness, the fitted value of the hydrogen column density was $4.7\times10^{20} \rm{cm}^{-2}$. Given the low statistics of the BAL sample, it is not possible to fit an independent column density reliably, so we cannot assess the origin of the X-ray weakness in this sample. We instead refer the reader to the detailed studies of a few BALs, quoted above.

%%%

\section{Conclusions}
In this work we present the first, to the best of our knowledge, systematic comparison of amplitude of the flux variability in BAL and non-BAL AGNs. To this end, we have measured the variance from optical ZTF light curves in the observer frame $g$ and $r$ bands, for a sample of 1031 BAL AGNs and 6624 non-BAL AGNs. All objects have $1.57<z<2$, an  apparent $r$ band magnitude in the range of [18.5,19.8], estimated \mbh\ in the range $7.5<\log $\mbh$<10.5$, and \redd\ within $-2<\log R_{Edd}<1$. All \mbh\ were estimated with the single-epoch method using the Mg\,\textsc{ii} 2798 \AA\ line. 
The main results from our analysis, presented in Section~4 are the following:

   \begin{enumerate}
      \item We estimated the variability amplitude of the BAL and non-BAL samples by measuring the normalized excess variance ($\sigma_{\mathrm{rms}}^{2}$). The median $\sigma_{\mathrm{rms}}^{2}$ is systematically lower for BAL quasars compared to non-BALs in both observed bands. Specifically, in the $g$ band, the median $\sigma_{\mathrm{rms}}^{2}$ is $(70.8 \pm 2.1)\times10^{-4}$ for BALs versus $(99.5 \pm 1.3)\times10^{-4}$ for non-BALs. Similarly, the $r$ band values are $(32.2 \pm 1.3)\times10^{-4}$ versus $(55.1 \pm 0.9)\times10^{-4}$, as illustrated in Fig. \ref{fig:xs}. Overall, the $g$ band light curves have higher excess variance than those in the $r$ band.
      \item Using DRW parameters to compare both populations at fixed \mbh\ and 
\redd\, we show that BAL QSOs systematically display suppressed variability 
compared to non-BAL QSOs. Specifically, BALs are roughly 30\% to 36\% less 
variable than non-BALs in the $g$ and $r$ bands, respectively. On the other hand, 
no significant or consistent differences are observed in their $\tau_{\mathrm{DRW}}$ 
values.
      \item Measuring $\sigma_{\mathrm{DRW}}$ at fixed $L_{\mathrm{bol}}$ and 
redshift reveals that in the $r$ band, BAL QSOs show consistently lower 
variability independent of whether the C\,\textsc{iii}] line falls within the 
filter coverage. In the $g$ band, however, this reduced variability mostly 
disappears, but only within the highest redshift bin at the highest luminosities -- where the C\,\textsc{iv} trough is strongly present. 
   \end{enumerate}

These results indicate that the suppression of variability in BAL quasars is a fundamental property of their continuum emission. While we find that absorption line responsivity in the $g$ band -- specifically from the \ion{C}{iv} BAL -- introduces an additional source of variability, its effect is localized. This line-driven variability is strong enough to counter the underlying continuum suppression precisely in the highest-redshift, high-luminosity bin where the trough dominates the filter window, explaining why the intrinsic continuum suppression is effectively masked there. However, across the rest of the parameter space, it is insufficient to overcome the intrinsic continuum suppression. Furthermore, we find that contributions from the Fe~\textsc{ii} pseudo-continuum in the $r$ band is too weak to account for the reported differences between $r$ and $g$ bands, and too similar between BALs and non-BALs to account for the differences between the samples.

To explain this inherent continuum stability in BAL quasars, we explore three potential scenarios:
\begin{itemize}
\item Biased $M_{\mathrm{BH}}$ estimates: A systematic overestimation of virial masses in BALs, but not in the non-BALs, potentially due to non-virial components in the Mg~\textsc{ii} line, would lead to higher-than-estimated accretion rates. While our analysis shows that a mass correction of $\sim 0.6$~dex could reconcile the observed differences in variability amplitude, this scenario is somewhat challenged by our $\tau_{\mathrm{DRW}}$ measurements, which show no corresponding timescale offsets between the populations. Nevertheless, we cannot definitively rule out a mass bias, as subtle timescale shifts might currently be obscured by the statistical uncertainties, which require larger samples to resolve. Additionally, the values of $\tau_{\mathrm{DRW}}$ obtained are only about 4.5 times shorter than the length of the light curves in the rest frame, so these might be biased due to the limited length of our data.

\item Nuclear obscuration: The lower variability in BAL quasars may be linked to the presence of dust within the outflows. This dusty material, which is consistent with the redder colours observed in the BAL population, could effectively mask or smooth the rapid fluctuations of the central engine, leading to a more stable continuum compared to non-BAL sources.

\item X-ray weakness: Our cross-match with eROSITA-DE data reveals a clear deficit in X-ray detections for BALs ($1.6\%$ vs. $21\%$ for non-BALs) and stacked eROSITA spectra shows that, on average, the BAL sample is about 8 times X-ray weaker than the non-BAL sample. In the framework of X-ray reprocessing, a weaker incident X-ray flux would naturally result in a less variable UV/optical accretion disc, providing a compelling physical basis for our findings. We note that for this scenario to work the BALs should be intrinsically X-ray weak, or alternatively, have their coronal emission blocked from the accretion disc as well as from the observer. 
\end{itemize}

\begin{acknowledgements}
Authors thank for the helpful comments and suggestions from the anonymous referee, which helped to improve the robustness of the analysis and the clarity of the presentation.
We acknowledge the Millennium Science Initiative Programs NCN$2023\_002$ (AA, PA, MLM, BM), FONDECYT Regular 1241422 (PA), CAV, CIDI N. 21 U. de Valparaíso, Chile (PA) and the China-Chile Joint Research Fund (CCJRF2310)(MLM). Based on observations obtained with the Samuel Oschin Telescope 48-inch and the 60-inch Telescope at the Palomar Observatory as part of the Zwicky Transient Facility project. ZTF is supported by the National Science Foundation under Grants No. AST-1440341 and AST-2034437 and a collaboration including current partners Caltech, IPAC, the Weizmann Institute for Science, the Oskar Klein Center at Stockholm University, the University of Maryland, Deutsches Elektronen-Synchrotron and Humboldt University, the TANGO Consortium of Taiwan, the University of Wisconsin at Milwaukee, Trinity College Dublin, Lawrence Livermore National Laboratories, IN2P3, University of Warwick, Ruhr University Bochum, Northwestern University and former partners the University of Washington, Los Alamos National Laboratories, and Lawrence Berkeley National Laboratories. Operations are conducted by COO, IPAC, and UW.
\end{acknowledgements}

\bibliographystyle{aa} 
\bibliography{ref}

\begin{appendix}

\section{Median properties of the BAL and non-BAL samples}
\label{sec:appendix}

\begin{table*}
\centering
\caption{Median physical properties of the paired BAL and non-BAL sub-samples across the $3 \times 3$ grid in \mbh\ and \redd .}
\label{tab:bin_stats}
\begin{tabular}{cc rcccc rcccc}
\toprule
\multicolumn{2}{c}{Bin Ranges} & \multicolumn{5}{c}{non-BAL Sample} & \multicolumn{5}{c}{BAL Sample} \\
\cmidrule(r){1-2} \cmidrule(lr){3-7} \cmidrule(l){8-12}
$\log \lambda_{\mathrm{Edd}}$ & $\log M_{\mathrm{BH}}$ & $N$ & $z_{\mathrm{med}}$ & $\sigma_z$ & $\log L_{\mathrm{bol, med}}$ & $\sigma_{\log L}$ & $N$ & $z_{\mathrm{med}}$ & $\sigma_z$ & $\log L_{\mathrm{bol, med}}$ & $\sigma_{\log L}$ \\
\midrule
$[-1.20, -0.90)$ & $[9.10, 9.40)$ &  481 & 1.710 & 0.007 & 46.424 & 0.005 &   71 & 1.709 & 0.022 & 46.440 & 0.014 \\
$[-1.20, -0.90)$ & $[9.40, 9.70)$ &  706 & 1.804 & 0.006 & 46.635 & 0.005 &  135 & 1.832 & 0.015 & 46.635 & 0.012 \\
\addlinespace
$[-0.90, -0.60)$ & $[8.80, 9.10)$ &  571 & 1.698 & 0.005 & 46.423 & 0.005 &   64 & 1.722 & 0.022 & 46.450 & 0.011 \\
$[-0.90, -0.60)$ & $[9.10, 9.40)$ & 1832 & 1.783 & 0.005 & 46.616 & 0.004 &  259 & 1.812 & 0.011 & 46.626 & 0.008 \\
$[-0.90, -0.60)$ & $[9.40, 9.70)$ &  613 & 1.852 & 0.006 & 46.815 & 0.005 &  115 & 1.865 & 0.015 & 46.792 & 0.010 \\
\addlinespace
$[-0.60, -0.30)$ & $[8.80, 9.10)$ &  887 & 1.769 & 0.012 & 46.619 & 0.004 &  123 & 1.791 & 0.027 & 46.613 & 0.013 \\
$[-0.60, -0.30)$ & $[9.10, 9.40)$ &  547 & 1.845 & 0.007 & 46.803 & 0.005 &   71 & 1.834 & 0.019 & 46.781 & 0.024 \\
\bottomrule
\end{tabular}
\end{table*}

\end{appendix}

\end{document}